\documentstyle[12pt]{article}

\newcommand{\be}{\begin{equation}}
\newcommand{\ee}{\end{equation}}
\newcommand{\bea}{\begin{eqnarray}}
\newcommand{\eea}{\end{eqnarray}}
\newcommand{\nn}{\nonumber \\}
\newcommand{\p}[1]{(\ref{#1})}
\newcommand{\ba}{\begin{array}}
\newcommand{\ea}{\end{array}}
\newcommand{\vs}[1]{\vspace{#1 mm}}

\renewcommand{\a}{\alpha}
\renewcommand{\b}{\beta}
\renewcommand{\c}{\gamma}
\renewcommand{\d}{\delta}
\newcommand{\e}{\epsilon}

\def\bbox{{\,\lower0.9pt\vbox{\hrule \hbox{\vrule height 0.2 cm
\hskip 0.2 cm \vrule height 0.2 cm}\hrule}\,}}
\newcommand{\dsl}{\pa \kern-0.5em /}

\newcommand{\pa}{\partial}

\font\mybb=msbm10 at 12pt
\def\bb#1{\hbox{\mybb#1}}

\def\bR {\bb{R}}

\def\a{\alpha}
\def\b{\beta}

\def\g{\gamma}

\def\d{\delta}

\def\s{\sigma}

\def\ds{|\partial s|^2}
\def\dsb{|\bar \partial s|^2}
\def\bz{{\bar z}}

\def\ie{{\it i.e.,}\ }
\def\eg{{\it e.g.,}\ }

\begin{document}

\topmargin 0pt
\oddsidemargin 5mm

\renewcommand{\thefootnote}{\fnsymbol{footnote}}
\begin{titlepage}

\setcounter{page}{0}

\rightline{\small \hfill hep-th/9811024}
\vskip -.75em\rightline{\small \hfill QMW-PH-98-40}
\vskip -.75em\rightline{\small \hfill KCL-TH-98-43}

\vs{15}
\begin{center}
{\Large Supersymmetric Fivebrane Solitons}
\vs{10}

{\large
Jerome P. Gauntlett$^1$, Neil D. Lambert$^2$ and Peter C. West$^2$
 } \\
\vs{5}
${}^1${\em Department of Physics\\
       Queen Mary and Westfield College\\
       University of London\\
       Mile End Road\\
       London E1 4NS, UK}\\[2em]
${}^2${\em Department of Mathematics\\
King's College\\
University of London\\
The Strand\\
London WC2R 2LS, UK}
\end{center}
\vs{7}
\centerline{{\bf Abstract}}
We study the conditions for the fivebrane worldvolume theory
in D=11 to admit supersymmetric solitons with non-vanishing self-dual 
three-form. We construct some
new soliton solutions consisting of ``superpositions'' of 
calibrated surfaces, self-dual strings and instantons.

\end{titlepage}
\newpage
\renewcommand{\thefootnote}{\arabic{footnote}}
\setcounter{footnote}{0}

\section{Introduction}

An interesting feature of soliton solutions of brane worldvolume 
theories is that they contain their own spacetime 
interpretation \cite{cm,g,hlw}. 
A simple example is the
self-dual string soliton of \cite{hlw} which has the spacetime interpretation
of a membrane ending on a fivebrane. Since the corresponding supergravity
solutions are typically not fully localised (see \cite{gauntlett} for a review)
the worldvolume theory provides a sensitive tool 
to study the properties of brane intersections.
In particular they have a number of applications 
within string theory and also Yang-Mills theory.

In \cite{glw} a study was initiated of
interpreting general static spacetime configurations of intersecting 
branes (\eg supergravity
solutions) in terms of solitons on the worldvolume. In that paper we only
considered configurations of intersecting fivebranes
which correspond to solitons  on the worldvolume 
with vanishing self-dual three-form. These
configurations can be interpreted as a single static fivebrane with a 
non-planar worldvolume. Furthermore one finds that the 
supersymmetric worldvolume
solitons correspond to fivebranes whose spatial worldvolumes form 
calibrated surfaces \cite{bbs,oog,glw}\footnote{For another
connection between calibrations and intersecting branes, see 
\cite{gp,afs,afs2,afss,jose}.}. In particular we showed that the differential
equations for calibrated surfaces derived in~\cite{hl} 
are equivalent to the preservation of some
of the worldvolume supersymmetries.

In this paper we continue the study of supersymmetric 
fivebrane solitons 
by considering configurations with non-vanishing
three-form. 
{}From a mathematical point of view 
this is a natural generalisation of calibrations.  From a physical
point of view, one expects to find supersymmetric solitons that include 
configurations corresponding to intersecting fivebranes and membranes and
M-waves in spacetime.
Here we shall explore several interesting cases, all in D=11 Minkowski
space, leaving further analysis
and applications to future work.

The solitons constructed here may be thought of as arising 
from three ``building
blocks''. The first of these 
has the
spacetime interpretation as a membrane intersecting a fivebrane which
we may denote as
\be
\matrix{M5:&1&2&3&4&5&\cr
        M2:& & & & &5&6\cr}
\ee
On the worldvolume of the fivebrane 
this solution has a single scalar $X^6$ active, depending
on the four coordinates $\s^m =\{\s^1,\s^2,\s^3,\s^4\}$. It also has 
non-vanishing
$H_{05m}$ and, as a consequence of the self-duality 
constraint imposed on $H$, 
$H_{mnp}$. The resulting 
configuration on the worldvolume is a self-dual string soliton parallel to
the $\s^5$ direction \cite{hlw}.

A second building block solution which we will use can be thought of
as an M-wave intersecting a fivebrane
\be\label{blue}
\matrix{M5:&1&2&3&4&5&\cr
        MW:& & & & &5&\cr}
\ee
This solution has no scalars active. We will see below that the
three-form for this configuration has non-vanishing
$H_{0mn}$ and $H_{5mn}$, where $m=1,2,3,4$. In fact 
$H_{5mn}$ is self-dual as a two-form and gives rise to a string-like
soliton in the $\s^5$ direction.
By dimensional reduction we simply obtain an instanton in the
worldvolume Dirac-Born-Infeld (DBI) theory of a D4-brane.
This corresponds to a D0-brane in a D4-brane~\cite{douglas}
which is the configuration that 
is obtained by reducing \p{blue} along the $5$ direction.
In contrast to the self-dual string above, these strings
do not carry any charge with respect to the field $H$ and we therefore
also refer to these instanton solutions as neutral strings.

The final basic building block solutions which we will use are the
calibrated surfaces corresponding to intersecting fivebranes. 
The simplest example is provided by two intersecting
fivebranes
\be
\matrix{M5:&1&2&3&4&5& & & \cr
        M5:& & &3&4&5& &7&8\cr}
\label{riemann}
\ee
The corresponding worldvolume solution 
has two scalars $X^7,X^8$ active that depend on
the two worldvolume coordinates $\s^1,\s^2$. The 
general solution can be viewed
as a single fivebrane wrapped around a calibrated 
surface. In this case the relevant calibration is  K\"ahler so that 
the embedding $X^7(\s^1,\s^2), X^8(\s^1,\s^2)$ defines a  Riemann surface.
However our analysis will include considerably more complicated 
intersections such as
\be
\matrix{
           M5:&1&2&3&4&5& & & & & &\cr
           M5:& & &3&4&5& &7&8& & &\cr
           M5:& &2& &4&5& &7& &9& &\cr
           M5:&1&2& & &5& &7&8& & &\cr
           M5:&1& &3& &5& &7& &9& &\cr
           M5:& &2&3& &5& & &8&9& &\cr
           M5:&1& & &4&5& & &8&9& &\cr
           M5:& & & & &5& &7&8&9&10&\cr
           M5:& &2&3& &5& &7& & &10&\cr
           M5:& & &3&4&5& & & &9&10&\cr
           M5:& &2& &4&5& & &8& &10&\cr
           M5:&1& &3& &5& & &8& &10&\cr
           M5:&1& & &4&5& &7& & &10&\cr
           M5:&1&2& & &5& & & &9&10&\cr}
\label{cayle}
\ee
The corresponding worldvolume solitons have 
four scalars active depending on four worldvolume coordinates and
it can be viewed as a single fivebrane wrapped on a 
Cayley four-fold of ${\bf R}^8$~\cite{glw}.

In \cite{glw} we used orthogonal configurations of fivebranes 
to motivate the search for supersymmetric solitons.  This point of view 
suggests what
scalar fields should be active and what projections to impose on the
spinor parameters. The logic in this paper is similar. In particular
we note that the orthogonal configurations above can be combined (usually
breaking more supersymmetry) and this suggests that the 
corresponding worldvolume
solitons can similarly be ``superposed''.
Adding a membrane in the 5,6 directions to \p{riemann} and \p{cayle}
and other configurations in \cite{glw} suggests that we can add a 
self-dual string to the corresponding calibrated surface. Similarly 
adding a wave in the 5 direction suggests that we can 
add instantons to the calibrated
surfaces.
We shall see that this is indeed the case and that moreover
it is possible to combine all three. 

With this in mind here 
we will only consider calibrated surfaces $M$ of dimension $n$
with $n\le 4$.
\ie the fivebrane worldvolume has the form
${\bf R}^{6-n}\times M$. These have the feature that there is at least
one common flat  direction that is an isometry 
(\eg 5 in \p{cayle}) 
and one overall transverse
direction (\eg 6 in \p{cayle}) (of the cases considered in \cite{glw} only
the five-dimensional special
Lagrangian manifolds are excluded). 

To help illustrate this point  let us
list the spacetime configurations for the simplest example of a 
calibration in this class \p{riemann}.
For this case the 
corresponding soliton solutions that we
discuss  can be pictured as
\be
\matrix{M5:&1&2&3&4&5& & & \cr
        M5:& & &3&4&5& &7&8\cr
        M2:& & & & &5&6& & \cr}
\label{simple}
\ee 
where we have added a membrane, 
\be
\matrix{M5:&1&2&3&4&5& & & \cr
        M5:& & &3&4&5& &7&8\cr
        MW:& & & & &5& & & \cr}
\label{instcal}
\ee
where we have added an M-wave, and finally
\be
\matrix{M5:&1&2&3&4&5& & & \cr
        M5:& & &3&4&5& &7&8\cr
        M2:& & & & &5&6& & \cr
        MW:& & & & &5& & & \cr}
\label{allcal}
\ee
where both have been added.

Another type of solution which we will consider, consists of adding
membranes intersecting at a point to a fivebrane
\be
\matrix{M5:&1&2&3&4&5& & & \cr
        M2:& & & &4& &6& &\cr
        M2:& & & & &5& &7& \cr}
\label{sdsds}
\ee
which corresponds to two intersecting self-dual strings.

The analysis in this paper is divided into two parts. 
The first part obtains
static soliton solutions using the Hamiltonian formalism. The general 
procedure  for finding supersymmetric states in this formalism  is described  
in the next section.
In section 2.1  we go on to  consider adding self-dual strings to calibrated 
surfaces.
In section 2.2 we discuss 
instanton configurations 
and  discuss their
interpretation as 
neutral strings. 
Next in section 2.3 we consider 
instantons on calibrated surfaces.
In section 2.4 we then consider 
superpositions of neutral strings and self-dual strings 
on a calibrated surface. In section 2.5 we will
describe the case of two orthogonally intersecting self-dual strings
\p{sdsds}, although in this case we
will not include the dependence of the fields on the 4,5 directions in
\p{sdsds}.

The second part of the paper focuses on obtaining 
soliton solutions using the manifestly 
covariant formalism. The general supersymmetry conditions for this 
formalism are derived in
section 3. 
In section 3.1 we consider time-dependent
supersymmetric states corresponding to travelling waves along  
a fivebrane wrapped on a calibrated surface. These
states simply correspond to a ripple in the ``shape'' of the calibrated 
surface,  travelling along a flat direction \ie 
the string intersection such as 5 in \p{cayle}.
In section 3.2 we repeat the construction of instantons on a calibrated
surface to 
obtain neutral strings, including a generalisation to non-static instantons. 
Section 3.3 considers self-dual
and neutral strings together on a flat fivebrane. 
In 
section 3.4  we reconsider two intersecting self-dual strings, this time
including the dependence on the 4,5 transverse directions in \p{sdsds}. 
Lastly we conclude in 
section 4 with some comments.

\section{Supersymmetry in the Hamiltonian Formalism}

Let us begin  by describing  the Lagrangian formulation of the 
fivebrane in D=11 Minkowski space~\cite{pst}.
The bosonic variables are scalars $X^\mu$, $\mu=0,1,2,...,10$, 
a closed three-form $H_{ijk}$, $i=0,1,2,...,5$, and an auxiliary scalar 
field $a$. The role of $a$ is to impose a non-trivial self-duality
constraint on $H$.
The supersymmetry and $\kappa$-symmetry transformations
of the fermions are given by
\be
\label{one}
\delta\theta={1\over 2} (1+\Gamma)\kappa + \epsilon
\ee
where $\epsilon$ is a constant 32 component $D$=11 spinor.
The matrix $\Gamma$  is given by
\bea
\Gamma= {1\over {\sqrt{-{\rm det}(g+\tilde H)}}}
\bigl[&{}&{1\over 5!}{1\over (\pa a)^2}(\pa_i  a\Gamma^i)\Gamma_{i_1\dots i_5} 
\epsilon^{i_1\dots i_5j}\pa_j a+\pa_i a \Gamma^i \Gamma_j t^j \nn
&-& {{\sqrt {-g}}\over 2 (-(\pa a)^2)^{1\over 2}}
\pa_i a\Gamma^i\Gamma_{jk} \tilde H^{jk}
\bigr]\ ,
\eea
where
\bea
g_{ij} &=&\partial_iX^{\mu}\partial_jX^{\nu}\eta_{\mu\nu}\ ,\nn
\tilde H^{ij} &=& {1\over (-(\pa a)^2)^{1\over 
2}}(*H)^{ijk} \pa_k a\ ,\nn
t^i &=& {1\over 8 (\pa a)^2} \varepsilon^{ij_1j_2k_1k_2l} 
\tilde H_{j_1j_2}\tilde H_{k_1k_2}\pa_l a\ ,\nn
\Gamma_i&=&\pa_i X^\mu \gamma_\mu\ .
\eea
Here $g_{ij}$ is the pull back of the eleven-dimensional
flat metric,  $*H$ is the Hodge dual of $H$ with respect to $g$ and
$\gamma^\mu$ are flat $D$=11 gamma-matrices. 
Note that $\Gamma^2=1$ and hence ${1\over 2}(1\pm \Gamma)$ are projection 
operators.
{}For a bosonic configuration the corresponding variation
of the bosonic fields $X^\mu$ and $H$ automatically vanish.
Thus bosonic configurations will preserve some
supersymmetry if and only if \p{one} vanishes. This is equivalent to the 
condition\footnote{This form for the preservation of
supersymmetry of was first discussed
in~\cite{bbs} and was subsequently considered in \cite{oog,bkop,glw,gp}.}
\be
(1-\Gamma)\epsilon =0\ .
\ee

{}For  most of  this  paper we will study static configurations
for which  the Hamiltonian formalism is most convenient. We will work in 
static gauge, $X^i=\sigma^i$ and 
in addition 
choose the gauge $a=t$. 
In this case $t^0=\tilde H^{0a}=0$, where $a=1,\dots,5$ is a spatial
index. Furthermore, one can check that $\Gamma^\dag =\Gamma$.
Using this we have
\bea
||(1-\Gamma)\epsilon||
=\epsilon^\dagger(1-\Gamma^\dag)(1-\Gamma)\epsilon&\ge0&\nn
\Longleftrightarrow\epsilon^\dag(1-\Gamma)\epsilon&\ge0&\ .
\eea
Choosing a spinor satisfying $\epsilon^\dag \epsilon=1$ we thus deduce
the Bogomol'nyi bound on any static configuration
\be
1\ge \e^\dag \Gamma\e\ ,
\ee
with equality for supersymmetric configurations.
The bound
can be rewritten in the form
\be\label{bound}
{\sqrt{{\rm det}(g+\tilde H)}}\ge
\epsilon^\dag
\Gamma^0\left[\Gamma_a t^a - {{\sqrt {g}}\over 2 }
\Gamma_{ab} \tilde H^{ab}
+{1\over 5!}\Gamma_{a_1\dots a_5} \epsilon^{a_1\dots a_5}
\right]\e\ ,
\ee
where
\bea
\tilde H^{ab} &=& {1\over 3!}
{1\over \sqrt{g}}\varepsilon^{abc_1c_2c_3} 
H_{c_1c_2c_3} \ ,
\nn
t_f &=& {1\over 4!} \varepsilon^{abcde} H_{abc}H_{def}\ ,
\eea
and we use the convention that $\varepsilon^{12345}=1$ (and hence 
$\varepsilon_{12345}={\rm det}g$). 

One expects that this condition should provide a bound on the energy.
The energy functional of static configurations in static gauge
are given by~\cite{bergtown,ggt}
\be\label{energy}
{\cal E}^2 = \det (g_{ab} + \tilde H_{ab}) +t_a t_b m^{ab}\ ,
\ee
where
\be
\label{emmeqn}
m^{ab} = g^{aa'}g^{bb'} \big[ \partial_{a'} {\bf X} \cdot \partial_{b'}{\bf X}
+ (\partial_{a'}{\bf X}\cdot \partial_{c}{\bf X})\delta^{cd}
(\partial_{b'}{\bf X}\cdot \partial_{d}{\bf X})\big]\ .
\ee
Noting that $m^{ab}=\pa_a{\bf X}\cdot \pa_c {\bf X} g^{cb}$
we see that $m^{ab}$ is a positive definite matrix and hence we
deduce that the energy for all static configurations of
the fivebrane is also bounded by the right hand side of \p{bound}.
{}For configurations where $t_at_b m^{ab}=0$, the condition for the 
bound on the energy being saturated is the same as the
condition for preservation of supersymmetry (\ie saturation of \p{bound}).
In the rest of the paper we will mainly consider this case. 
However, since
the preservation of some supersymmetry generally implies that the
energy is minimised it should be possible to derive a better bound on
the energy such that when it is saturated it is equivalent to preservation of 
supersymmetry
even when $t_at_b m^{ab} \ne 0$, but we shall not pursue this here.

\subsection{Self-Dual Strings on Calibrated Surfaces}

Let us now begin our construction of  supersymmetric solutions by 
superimposing a self-dual string on a calibrated surface. 
{}First consider a configuration of fivebranes only with at least one overall
string intersection,
\eg \p{riemann} or \p{cayle}.
On the first fivebrane worldvolume these correspond to 
configurations 
with the scalars $X^I, I=7, \dots, 10$ being non-trivial functions
of the world-volume coordinates $\sigma^\alpha, \alpha = 1,\dots,4$. 
It was shown in \cite{glw} that 
configurations that preserve supersymmetry correspond to calibrated fivebrane 
world-volumes, the
calibration depending on the particular case being considered.
Note that if all of the scalars are excited then the calibrated surface $M$ is
four dimensional and the spatial part of the fivebrane worldvolume takes the 
form
$\bR\times M$. In general if not all of the four scalars are excited then this 
becomes
$\bR^{5-n}\times M$ where the calibrated surface $M$
is now $n$ dimensional with $n=2,3,4$. 

Since we can add a membrane to these fivebrane 
intersections while preserving supersymmetry,
we expect to find supersymmetric 
self dual strings superposed on the corresponding calibrated
fivebrane worldvolume.
A concrete example is given by adding a membrane to \p{riemann} as pictured
in \p{simple}. 
The two fivebranes correspond to a
spatial worldvolume of the fivebrane given by $\bR^3\times \Sigma$ 
where $\Sigma$ is a two dimensional Riemann surface lying in the $\s^3,\s^4$ 
directions. We then expect to be able to add a 
self-dual string along the $\s^5$ direction.

Let us now consider the conditions required for the self-dual string to
preserve supersymmetry and solve the equations of motion.
We first write the  scalar coordinate not appearing in the calibrated surface as
\be
X^6\equiv X\ ,
\ee
and we will demand $\pa_{0,5} X=\pa_{0,5} X^I=0$. 
The induced spatial worldvolume metric can be written as
\be
g_{\alpha\b} =\bar g_{\a\b} + \pa_\a X \pa_\b X\ ,\quad
g_{\alpha 5} =0\ ,\quad
g_{55} =1\ ,
\ee
where 
\be
\bar g_{\a\b} = \delta_{\a\b}+\pa_\a X^{I} \pa_\b X^{J}\delta_{IJ}\ .
\ee
Our ansatz for a supersymmetric solution will include taking 
taking $\bar g$ to be the induced metric on $R^{4-n}\times M$ where $M$ is
an $n$ dimensional calibrated 
surface.
As discussed in detail in \cite{glw} this by itself leads
to a supersymmetric solution where both the number
of supersymmetries preserved and the type of calibration
are determined by certain projections on the supersymmetry
parameters. These projections are the same as those for the corresponding
spacetime configurations of orthogonally intersecting fivebranes.
Here we will need that the condition for supersymmetry to be preserved is given
by
\be
\label{calcond}
\g^0\bar\Gamma_{1234}\g_5\e ={\sqrt {\bar g}}\e\ ,
\ee
where $\bar\Gamma_\a=\g_\a+\pa_\a X^{I}\g_{I}$. By multiplying both
sides by $\e^\dag$ and imposing the
relevant projection operators we obtain the calibration
$\Omega$ via
\be
\Omega_{\a_1\a_2\a_3\a_4}=\e^\dag\g^0\bar\Gamma_{\a_1\a_2\a_3\a_4}\g_5\e\ .
\ee

The ansatz for including a membrane is obtained by ``superposing''
this with the ansatz for the self dual string. Specifically, we take
\be
\tilde H^{\a 5} =\pm{{\sqrt {\bar g}}\over \sqrt g}\bar g^{\a\b}\pa_\b X\ ,
\label{sds}
\ee
or equivalently
\be
H=\pm\bar * d X\ ,
\ee
where $\bar *$ is the Hodge dual with respect to the metric $\bar g$. 
Since $H$ is closed,
$X$ must be a harmonic function\footnote{We note that related solutions were 
argued to 
solve the DBI equations of motion in \cite{gp}.}
on the 
geometry specified by $\bar g$.
Note that in general this is not quite the same as saying that
it is harmonic on the calibrated surface $M$. 
Recall that if not all of the scalars $X^I$ are excited then the
geometry determined by $\bar g$ is actually $\bR^{4-n}\times M$.

To further illustrate this
point consider the configuration \p{simple}, \ie $n$=2.
If we let $\hat g$ specify the metric of the calibrated surface,
in this case the Riemann surface $\Sigma$, and
if we let $i=1,2$ and $a=3,4$ (for here only), then $X$
satisfies
\be
\pa_a\pa_a X
+{1\over{\sqrt {\hat g}}} \pa_i ({\sqrt {\hat g}} \hat g^{ij}\pa_j) X
=0 \ .
\label{moregen}
\ee
If $\pa_a X=0$ then this is equivalent to $X$ being harmonic
on the calibrated surface $\hat g$. Solutions are provided by the
real parts of holomorphic functions, but it should be noted
that the string is then delocalised
in the $3,4$ directions. A localised string is obtained by
solving the more general equation \p{moregen}.

To analyse the conditions for preserved supersymmetry we first observe that
$t^a=0$ and hence from \p{energy} that ${\cal E}=\sqrt{\det (g_{ab} + \tilde 
H_{ab})}$. Noting that 
\be\label{lemdetone} 
\det g= \det \bar g(1+(\overline{\pa X})^2)\ ,
\ee
where we have introduced the notation
$(\overline{\pa X})^2=\bar g^{\a\b}\pa_\a X \pa_\b X$,
we can use the
the expansion
\be
\label{lemdet}
\det (g_{ab} + \tilde H_{ab})=
\det g(1 + {1\over 2} \tilde H^2) + t^2
\ee
to show that ${\cal E}^2$ is a perfect square with
\be\label{eone}
{\cal E}=\sqrt{\det (g_{ab} + \tilde 
H_{ab})}={\sqrt {\bar g}}(1+(\overline{\pa X})^2)\ .
\ee
This should be compared to the right-hand side of \p{bound}.
Using the fact that $\Gamma_a = \bar \Gamma_a + \partial_a X\gamma_6$,
we have 
\be
\Gamma_{12345}=\bar\Gamma_{1234}\g_5 +\pa_\a X \bar 
g^{\a\b}\bar\Gamma_\b\bar\Gamma_{1234}\g_{56}\ ,
\ee
and
\be
-{{\sqrt g}\over 2}\Gamma^{ab} \tilde H_{ab}
=\mp{\sqrt{\bar g}}\bar g^{\a\b}\pa_\b 
X\left(\pa_\a X \g_6\g_5 +\bar \Gamma_\a 
\gamma_5\right)\ .
\ee
If in addition to demanding \p{calcond} we also insist that 
\be
\label{memproj}
\g^{056}\e =\pm\e\ ,
\ee
which is the supersymmetry condition for a self-dual string without
the calibrated surface, then the right side of the supersymmetry 
condition precisely becomes \p{eone}.
Thus the ansatz preserves
supersymmetry.
To determine the amount of supersymmetry 
one needs to consider the specific calibration. In general
one expects that the addition of the membrane should
break a further half of the supersymmetry. However, in some
cases the projection is a consequence of the projections imposed
by the surface being calibrated. 
{}From a spacetime point of view,
this corresponds to configurations of fivebranes where the
membrane can be ``added for free'' (see section 2 of \cite{glw} for examples).
To fully specify the solution, one needs to find harmonic functions
on calibrated surfaces (or solve the analogue of \p{moregen}.

\subsection{Instantons and Neutral Strings}

It is well known that a D0-brane bound to a D4-brane
manifests itself as an instanton in the D4-brane world
volume theory~\cite{douglas}. 
{}For a single D4-brane it was shown in \cite{ggt} that Abelian
instantons saturate a Bogomol'nyi bound in the full DBI theory. 
It was further argued in \cite{ggt} that
the bound should also hold in the non-Abelian theory,
corresponding to a collection of coincident D4-branes, and this was 
confirmed in \cite{brecher}. 

As a spacetime configuration a D0-brane intersecting a D4-brane
uplifts to eleven dimensions as a gravitational pp-wave or M-wave intersecting a
fivebrane according to the pattern \p{blue}.
We thus expect this configuration to be 
realised on the fivebrane worldvolume theory as an ``instanton string''.
We shall confirm this and then show that it can be
superposed with calibrated surfaces and self-dual strings in the following
subsections.

To construct the instanton string, 
we set all of the scalar fields to zero, and hence the induced metric
is flat, $g_{ab}=\delta_{ab}$. The ansatz for the $H$-field is
taken to be
\be\label{hinst}
H_{5\a\b}= F_{\a\b}\ ,
\ee
where $F=dA$ with $F$ an (anti-) self-dual field strength. 
We then have
\be
t_\a=0\ ,\qquad
t_5=\pm{1\over 4} F^2\ ,
\ee
and hence $t_a t_b m^{ab} =0$. As in \cite{ggt} 
${\rm det}(\delta +F)$ is a perfect square and
the energy is given by
\be
{\cal E} ={\sqrt{{\rm det}(g+\tilde H)}}= 1+{1\over 4} F^2\ .
\ee
{}For this configuration to preserve supersymmetry, this should be 
equal to the right hand side of \p{bound}. 
Substituting the ansatz, the latter is given by
\be\label{here}
\epsilon^\dagger \gamma^0[\gamma_5 t_5 +\gamma_{12345} 
\mp {1\over 2} \gamma^{\a\b} F_{\a\b}]\epsilon\ .
\ee
If we impose the projectors
\be
\gamma^{05}\epsilon = \pm\epsilon\ ,\qquad
\gamma^{012345}\epsilon =\epsilon\ ,
\ee
we see that the last term in \p{here} vanishes
and the Bogomol'nyi bound is saturated. The configuration breaks one half
of the world-volume supersymmetry corresponding to one quarter of the
spacetime supersymmetry.

Thus we find this configuration corresponds to (anti-) self-dual $U(1)$ gauge
fields on the space transverse to the string. Explicit solutions can be 
constructed
using a complex structure $J$ on $R^4$ via
\be\label{sol}
A_\a={J_\a}^\b\pa_\b \phi
\ee
where $\phi$ is a harmonic function on $R^4$ \cite{gp}. If the K\"ahler-form
is anti-self dual the field strength is self dual and vice-versa.
Note that, unlike the D4-brane, there is no known non-Abelian extension of the
classical fivebrane theory and thus the classical BPS solutions are
restricted to Abelian instantons.

It is worth emphasising that
these instanton strings differ from the self-dual strings of \cite{hlw}
in that they have vanishing $H$-charge
\be
Q = \int_{S^3}H = 0\ ,
\ee
where $S^3$ is the sphere at transverse spatial infinity. 
Thus we will also refer to these
instanton strings as ``neutral strings''\footnote{Neutral strings 
were discussed in the context of
the linearised approximation to the fivebrane theory compactified
on a torus in ~\cite{dvv}.}.

Note that the instanton can have an arbitrary 
dependence on $\sigma^5$, while still maintaining $dH=0$. In other
words the $\sigma^5$ direction specifies a one-dimensional path
in the moduli space of instantons. For example in the solutions \p{sol},
we can take 
\be\label{solt}
\phi= {f(\sigma^5)\over |\sigma^\a - h^\a(\sigma^5)|^2}
\ee
where $|\sigma|^2=\sum_{\a=1}^4\sigma^\a \sigma^\a$. In other words the
instanton is free to change its location, $h$, 
and also its ``amplitude", $f$, along the
length of the string. 

\subsection{Instantons on Calibrated Surfaces}

We now discuss an interesting generalisation by 
considering instantons
on calibrated surfaces.
As in the last section we let $X^I(\sigma^\a)$, $I=7,8,9,10$, $\a=1,2,3,4$
specify a calibrated surface with induced metric $\bar g$. 
The spatial
worldvolume metric therefore takes the form
\be
\quad g_{\a\b}=\bar g_{\a\b}\ ,\quad
g_{55} =1\ , \quad g_{5\a}=0\ .\ee
{}For a neutral string along the 5 direction,
the only non-zero component of $H_{abc}$ is as in \p{hinst}
\be\label{thiseq}
H_{5\a\b} = F_{\a\b}\ ,
\ee
and we take $F_{ab}$ to be
(anti-) self-dual on the calibrated surface specified by $\bar g$
\be
{1\over2\sqrt{\bar g}}\epsilon_{\a\b\g\d}F^{\g\d} = \pm F_{\a\b}\ ,
\ee
This leads to 
\be
{\tilde H}_{\a\b} = \pm F_{\a\b}\ ,\quad
t_5 = \pm{1\over 4}{\sqrt {\bar g}} F^2,\quad t_\a=0\ .
\ee
Since $t_a t_b m^{ab}=0$ and using \p{lemdet} we have
\be
\label{eninscal}
{\cal E}=\sqrt{\det(g + \tilde H)}=\sqrt{\bar g}(1 + 
{1\over4}F^2)\ .
\ee

To see that the solution preserves supersymmetry we want to
compare this to the right hand side of \p{bound}. We will
impose the projections for both the calibration and the instanton:
\be
\g^0\bar\Gamma_{1234}\g_5\e ={\sqrt {\bar g}}\e\ ,\quad
\gamma^{05}\e = \pm\e\ .
\label{ppsusy}
\ee
In~\cite{glw} examples of calibrations were given 
for which \p{ppsusy} followed from the intersecting fivebranes alone, \ie
one could add an M-wave `for free'. However one could always consider imposing
\p{ppsusy} as an addition constraint breaking another half of the 
supersymmetry.
A little algebra now shows that these conditions imply
\be\label{lemon}
{1\over2}\sqrt{\bar g}
\epsilon_{\a\b\g\d}\bar \Gamma^{\g\d}\e = \mp\bar \Gamma_{\a\b}\e\ .
\label{susyalt}
\ee
Substituting the ansatz into the right-hand side of \p{bound} we get
\be
\e^\dagger\Big[
\pm  t_5 + {1\over2}\sqrt{\bar g}\gamma_0\bar \Gamma^{\a\b}\tilde H_{\a\b} 
+ \sqrt{\bar g}
\Big]\e\ .
\label{redbound}
\ee
Upon imposing the projectors
on $\e$ we note that 
the second term vanishes by the self-duality of $F$ 
and we obtain $\pm t_5+{\sqrt {\bar g}}$  which is the same as 
\p{eninscal}.

The number of smooth instanton strings is thus given by the
number of (anti-) self-dual 
two forms on the calibrated
submanifold. An interesting case is for a Cayley 4-fold $M$.
It was shown in \cite{hl} that the K\"ahler two-forms $\omega_e$ 
associated to each of the complex structures of $R^8$, $J_e$, $e\in S^6$,
are anti-self
dual when restricted to $M$.

By dimensional reduction, the construction we have presented implies that
instantons on D4-brane worldvolumes wrapped around calibrated surfaces
are also supersymmetric. It seems likely that this will remain true
for the $U(N)$ non-Abelian extension of the DBI theory corresponding to
$N$ coincident D4-branes. It was shown in \cite{hl} that any Cayley four-fold
naturally admits anti-self dual $SU(2)$ Yang-Mills
fields. Such instantons correspond to D0-branes on two superposed
D4-branes wrapped around the Cayley four-fold. It 
would be interesting to investigate this in more detail.

\subsection{Self-Dual Strings and Neutral Strings on Calibrated Surfaces}

We now generalise the above cases by 
considering the addition of both self-dual and 
neutral strings (instantons) along the direction $\s^5$, to a
calibrated surface with metric $\bar g_{ab}$ in the 
$(\s^1,\s^2,\s^3,\s^4,X^7,X^8,X^9,X^{10})$ plane.
We will see that generically $m^{ab}t_at_b\ne 0$ and the resulting energy 
expression
exhibits new features. 

The ansatz consists of superposing a neutral string 
with the self-dual string on
a calibrated surfaces considered in section three. Thus the metric is given by
\be
g_{\alpha\b} =\bar g_{\a\b} + \pa_\a X \pa_\b X\ ,\quad
g_{\alpha 5} =0\ ,\quad
g_{55} =1\ ,
\ee
and the three-form is specified by
\bea
\tilde H^{\a\b} &=& \kappa {\sqrt{\bar g}\over \sqrt{g}}{\bar F}^{\a\b}\ .\nn
\tilde H^{\a 5} &=&
\lambda{{\sqrt {\bar g}}\over \sqrt g}\bar g^{\a\b}\pa_\b X\ ,
\eea
where $\kappa,\lambda$ are signs 
and again  $F$ is (anti-) self-dual in the $\bar g_{\a\b}$ metric:
\be
{1\over 2}{1\over \sqrt{\bar g}}
\e^{\a\b\g\d}F_{\g\d} = \kappa {\bar F}^{\a\b}\ ,
\ee
and the bar on $F$ denotes that we have raised the indices using
the metric $\bar g$ (note that we again have \p{thiseq}).
This leads to the following expressions for $t_a$
\be
t_5 = {\kappa\over 4}{\sqrt g}{\bar F}^2\ ,\quad t_\a 
= \kappa\lambda\sqrt{\bar g}{\bar g}^{\b\g}\partial_\b XF_{\a\g}\ . 
\ee
and the bar on $F^2$ indicates we have contracted the 
indices here with $\bar g$
rather than $g$. Using \p{lemdet} and \p{lemdetone} and the lemma
\be
{}F_{\a\b} \bar g^{\b\g} F_{\g\d}=-{1\over 4} {\bar g}_{\a\d}{\bar F}^2
\ee
we obtain 
\be
\label{blimp}
\sqrt{{\rm det}(g+\tilde H)} =\sqrt{\bar g}(1+({\overline{\partial X}})^2
+ {1\over 4}({\bar F}^2))
\ee

To check the preservation of supersymmetry we will impose the projections
\be
\g^0\bar\Gamma_{1234}\g_5\e ={\sqrt {\bar g}}\e\ ,\quad
\gamma^{05}\e=\kappa\e\ ,\quad \gamma^{056}\e = \lambda\e\ ,
\ee
As before the
$\Gamma$-matrices are given by $\Gamma_a = \bar \Gamma_a + \partial_a 
X\gamma_6$, where
$\bar \Gamma_a$ are the $\Gamma$-matrices for the calibrated surface. 
Using $\bar \Gamma_{\a\b}\tilde H^{\a\b}\e = 0$ and the facts
$\partial_a X t^a=0$,  
$t^\a =\lambda\sqrt{g}\partial_\b X\tilde H^{\a\b}$,  it is straightforward to
show that the right hand side of the supersymmetry condition \p{bound}
is the same as \p{blimp}.

{}For the generic case  the energy is no longer given by 
$\sqrt{{\rm det}(g+\tilde H)}$ since $t_at_bm^{ab}\ne0$. Instead we find
\bea
{\cal E}^2 &=& {\rm det}(g+\tilde H) + t_at_bm^{ab}\nn
&=&{\rm det} \bar g \big[1+({\overline {\partial X}})^2
+ {1\over 4}({\overline F}^2)\big]^2 
+ {\rm det}\bar g (\delta^{\a\b} - \bar g^{\a\b})
\partial_\g X \partial_\d X \bar {F_{\a}}^\g\bar {F_{\b}}^\d\ .\nn
\eea
Thus the energy contains a non-linear term arising from the interaction of
the self-dual string, neutral string and calibration. In the solutions
constructed before we saw that the strings behaved very much like
additional fields living on the calibrated surface. In those cases the
energy was precisely what one would expect for a scalar field or Maxwell
field living on the calibrated submanifold. However here we see that the energy
of the combined self-dual string, neutral string and calibration configuration
is greater than  simply the sum of the two string energies on a 
background calibrated surface.

\subsection{Two Intersecting Self-Dual Strings}

Consider two membranes orthogonally intersecting a fivebrane according
to the pattern \p{sdsds}. 
We expect that this should manifest itself as 
two intersecting self-dual strings in the 
fivebrane worldvolume theory. Here we will construct
such supersymmetric solutions. 
It should be 
pointed out in advance that the two self-dual strings will be 
delocalised in the directions tangent to the other string.
\ie in the $4,5$ directions on the fivebrane. 
In section  ten we will consider
the most general solution using the covariant formalism.

In these solutions two scalars in the fivebrane worldvolume theory are active,
$X^6,X^7$, which we will denote by $X$ and $Y$ respectively. These scalars are 
functions
of the worldvolume coordinates $\sigma^\alpha$, $\alpha=1,2,3$. 
We first observe that 
\be
{\rm det} g = 
1 + (\pa X)^2 + (\pa Y)^2 +(\pa X)^2(\pa Y)^2 - (\pa X\cdot  \pa  Y)^2\ ,
\ee
where \eg $(\pa X)^2=\pa X\cdot \pa X= \pa_\a X\pa_\a X$.
The ansatz for the three-form is given by
\be
\tilde H^{\a5}={\lambda\over {\sqrt g}}\pa_\a X, \qquad
\tilde H^{\a4}={\kappa\over {\sqrt g}}\pa_\a Y\ .
\ee
where $\kappa,\lambda$ are again signs. 
Closure of the three-form implies that $X$ and $Y$ are harmonic on
$\bR^3$:
\be
\pa^2 X=\pa^2 Y=0\ .
\ee

This ansatz implies that the only non-zero component of
$t^a$ is given by 
\be
t^\a=(\pa X\times \pa Y)^\a\ ,
\ee
where we are using the usual vector cross product
$(\pa X\times \pa Y)^\a = \epsilon_{\a\b\c}\pa_\b X\pa_\c Y$.
This implies that $t_a t_b m^{ab}$ vanishes and
from \p{emmeqn},\p{lemdet}
we find that
\be
{\cal E}=\sqrt {{\rm det}(g+\tilde H)}=\left[1+(\pa X)^2 + (\pa Y)^2 \right]\ .
\ee

The condition for preserved supersymmetry \p{bound}  now reads
\be 
\sqrt {{\rm det}(g+\tilde H)} = \epsilon^\dagger\Gamma^0\left[(\pa X\times \pa 
Y)^\a \Gamma_\a
+ \pa_\a X \Gamma_{\a5} - \pa_\a Y \Gamma_{\a4} + \Gamma_{12345}\right]
\epsilon\ .
\ee
The right hand side can be recast in the form
\bea
&\epsilon^\dagger\gamma^0\big[(\pa X \times \pa Y )\gamma_i(1+\gamma_{4567})
+\pa_i X \gamma_i (\gamma_{123456} -\lambda\gamma_5)
+\pa_i Y \gamma_i (\gamma_{123457} -\kappa\gamma_4)\nn
&+\gamma_{12345} -\lambda(\pa X)^2 \gamma_{65} - \kappa(\pa Y)^2 \gamma_{74}
-(\pa X\cdot \pa Y) (\lambda\gamma_{75} +\kappa\gamma_{64})\big]\epsilon\ .
\label{rhs}
\eea
If we now impose the supersymmetry projections
\be
\gamma^{056} \epsilon =\lambda\epsilon\ ,\qquad
\gamma^{047} \epsilon =\kappa\epsilon\ ,\qquad
\gamma^{012345}\epsilon =\epsilon\ ,
\label{projs}
\ee
then we find the first three terms in \p{rhs}
vanish and that the remaining terms combine precisely
to give $\sqrt {{\rm det}(g+\tilde H)}$.
Thus the configuration preserves $1/4$ of the worldvolume supersymmetry
which is what one expects for spacetime configurations that
preserves $1/8$ of the supersymmetry. Again we note that
the projections are exactly the same as in the spacetime configurations.

It is very likely that the
solution can be generalised to include more membranes. For example, if
we added a membrane in the 3,8 directions and demanded that
the active scalars are only dependent on the $\sigma^1,\sigma^2$ directions,
we expect the solution to be determined by three harmonic functions.
Note that in this case, such functions have logarithmic divergences.

\section{Supersymmetry in the Covariant Formalism}

Let us now turn our attention to obtaining supersymmetric 
solitons using the covariant 
formalism~\cite{hsw}. We will first present some details of the
general formalism, analogous to those in section two, before using it to
discuss some of the previous solutions, including some generalisations to
include time-dependence.
 
It is convenient to use a 
different choice of notation. We now let $a,b,c,..=0,1,2,3,4,5$ refer to
tangent indices and $m,n,p,...=0,1,2,3,4,5$ be world indices of the fivebrane
worldvolume. 
In this formalism one first introduces a self-dual three-form $h_{abc}$
and then constructs from it another three-form 
\be
H_{abc} = (m^{-1})_{a}^{\ d}h_{dbc} \ ,
\label{Hdef}
\ee
where
\be
m_{a}^{\ b} = \delta_{a}^{\ b} - 2 k_a^{\ b}\ ,\quad 
k_a^{\ b} = h_{acd}h^{bcd}\ .
\ee
Note that, as a consequence of self-duality, 
$h_{eab}h^{ecd} = \delta_{[a}^{\ [c}k_{b]}^{\ d]}$. 
{}From this one can see that 
$H$ is indeed totally anti-symmetric; $H_{abc} = H_{[abc]}$. 
The importance of the three-form $H_{abc}$ is that its equation of
motion simply states that $H$ is closed. It is useful to note that, because
of the self-duality of $h$, one has the formula~\cite{hsw}
\be
(m^{-1})_a^{\ b}  = Q^{-1}(\delta_a^{\ b} +2 k_a^{\ b})\ ,
\ee
where $Q = 1 - {2\over3}k_a^{\ b}k_b^{\ a}$.
Because the self-dual three-form $h$ is not closed it is not a convenient
field for discussing the physics. Instead one is primarily interested
in the closed, but not self-dual, three-form $H$. However $H$ does split up 
into its self-dual and anti-self-dual parts as~\cite{hsw}
\be
H^{+}_{abc} = Q^{-1}h_{abc}\ ,\quad 
H^{-}_{abc} = 2Q^{-1}k_{a}^{\ d}h_{bcd}\ .
\ee
Thus to obtain a formula for $h$ in terms of $H$ we need only evaluate the
function $Q$ in terms of $H$. 
To this end we write $H=H^+ + H^-$ and note that since $H^{+ 2}= H^{- 2}=0$
\bea
H^2 &=& 2H^{+}_{abc}H^{-\ abc}\nn
&=& 4 Q^{-2}h_{abc}k^{a}_{\ d}h^{dbc}\nn
&=& 4 Q^{-2}k_{a}^{\ d}k^{a}_{\ d}\nn
&=& 6 Q^{-2} (1-Q)\ .
\eea
The unique non-singular solution to this quadratic equation is
\be
Q = -{3\over H^2}\Bigg[1 - \sqrt{1 +{2\over3}H^2}\Bigg]\ .
\ee
Note that if $H$ is self-dual then $Q=1$. 
Thus whenever we see the three-form $h$ we may replace it by the identity 
\be
h_{abc}= -{3\over 2 H^2}\Bigg[1 - \sqrt{1 +{2\over3}H^2}\Bigg]
(H_{abc}+ {1\over 3!}\epsilon_{abcdef}H^{def})\  .
\ee
where $\epsilon^{012345}=1$.
The supersymmetry projector in this formalism has been described in~\cite{hsw}
and takes the form
\bea
\Gamma &=& 
-{1\over 6!} {1\over \sqrt{-g}}\epsilon^{m_1...m_6}\Gamma_{m_1...m_6} 
+ {1\over 3}h^{m_1m_2m_3}\Gamma_{m_1m_2m_3}\nn
&\equiv& \Gamma_{(0)}+ \Gamma_{(h)}\ ,
\eea
where $g$ is the determinant of the induced metric on the fivebrane and, 
as before, 
$\Gamma_m = \partial_m X^{\mu}\Gamma^{\underline a}$. Here 
$\Gamma^{\underline a}$, 
${\underline a}=0,1,2,...,10$, are flat eleven-dimensional 
$\Gamma$-matrices.  
Clearly when
the three-form is zero and we consider only static solutions, $\Gamma$ is
the same for both formalisms. 
One can show the following properties of $\Gamma$
\be
\Gamma^2 = 1\ ,\quad \Gamma_{(0)}^\dag = \Gamma_{(0)}\ .
\ee
However $\Gamma_{(h)}$, and hence $\Gamma$ also, are not Hermitian.
As before we may derive a bound from the inequality\footnote{Note that
the spinor notation is that of \cite{glw} and differs from section 2.}
\bea
||\epsilon(1-\Gamma)||^2 &=&
\epsilon(1-\Gamma)(1-\Gamma^\dag)\epsilon^\dag \ge 0\nn
\Longleftrightarrow\epsilon(1+{1\over2}\Gamma_{(h)}\Gamma_{(h)}^\dag)\e^\dag
&\ge& \e(\Gamma_{(0)} +\Gamma_{(h)} + \Gamma_{(h)}^\dag)\epsilon^\dag\ .
\eea
Unfortunately, although the left hand side is  manifestly 
positive definite, it is not
clear what its interpretation is. Physically one expects
that  $1+{1\over2}\Gamma_{(h)}\Gamma_{(h)}^\dag$ is related 
to the energy but we cannot check this as the Hamiltonian has not
yet been constructed in the covariant formalism variables.
In any case supersymmetric configurations 
satisfy $\epsilon(1-\Gamma)=0$ and so saturate this bound.

To use this formalism to construct supersymmetric solutions
we will gauge fix the
fivebrane.
The procedure for how to do this was 
discussed in detail in~\cite{glw} for the purely scalar case. 
Now we need to repeat this analysis
including the field $h_{abc}$. 
{}Firstly we gauge fix the fivebrane by choosing its worldvolume 
coordinates to be equal to the spacetime coordinates $\s^0=X^0,...,\s^5=X^5$.
This leaves the remaining five spacetime coordinates as scalar zero modes
on the fivebrane worldvolume $X^{a'}$, $a'=6,7,8,9,10$. 
The induced metric $g_{mn}$ on the fivebrane then takes the form
\bea
g_{mn} &=& \eta_{mn} + \partial_mX^{a'}\partial_n X^{a'}\nn
&=& \eta_{ab}e_{m}^{\ a}e_{n}^{\ b}\ .
\eea

Next we need to consider the spinors. The thirty-two component 
eleven-dimensional spinor indices $\underline \alpha$ 
naturally split up into two sixteen
component indices $\alpha=1,2,3,...,16$ and $\alpha'=1,2,3,...,16$.
The M-fivebrane preserves half of the thirty-two spacetime supersymmetries, 
which leaves sixteen supersymmetries $\e^\alpha$ in the worldvolume theory.
The sixteen spinor modes $\Theta^{\alpha'}$ become fermionic Goldstone fields  
in the worldvolume theory.
{}Furthermore these indices
can be reduced into $Spin(1,5)$ and $Sp(4)\cong SO(5)$ indices which 
we denote as $(\alpha,i)$ with $\alpha, i=1,2,3,4$. In particular
superscript indices decompose as $\alpha\rightarrow {}^{\alpha i}$ and
$\alpha'\rightarrow {}_{\alpha}^{\ i}$
and subscript indices decompose as $\alpha\rightarrow {}_{\alpha i}$ and
$\alpha'\rightarrow {}^{\alpha}_{\ i}$. 
{}For example the worldvolume supersymmetries  $\epsilon^{\alpha}$
and fermions $\Theta^{\alpha'}$ are written as $\epsilon^{\alpha i}$ 
and $\Theta_{\alpha}^{\ i}$ respectively. 
One can always tell in which
sense we mean a particular index,  depending on whether or not there are
$i,j$ indices present. For the sake of clarity we
will try to use as few spinor indices as possible without being ambiguous.
{}For a more complete discussion of the spinors  we refer
the reader to \cite{glw}.

{}Finally we must split up the eleven-dimensional $\Gamma$-matrices into a
six-dimensional form. For these we take
\bea
(\Gamma^{a'})_{\underline \alpha}^{\  \underline
\beta}=(\gamma^{a'})_{i}^{\ j}
\left (\matrix{  \delta_{\alpha}^{\beta}&0\cr
0&-\delta^{\alpha}_{\beta}\cr}\right)\ ,\quad
(\Gamma^a)_{\underline \alpha}^{\  \underline \beta}=\delta
_{i}^{j}\left (\matrix{ 0&(\gamma^a)_{\alpha \beta}\cr
(\tilde \gamma^a)^{\alpha \beta}&0\cr}\right)\ .
\label{gammadecomp}
\eea
Here $\gamma^{a'}$ are a set of five-dimensional Euclidean 
$\gamma$-matrices. The $\tilde\gamma$-matrices are simply 
related to the $\gamma$-matrices by
${\tilde \gamma}^a = {\gamma}^a $ for $a\ne 0$ and 
${\tilde \gamma}^0=-\gamma^0$. The matrices $\gamma^{a}$ must be chosen 
so that 
the $\Gamma^{a}$ satisfy a six-dimensional 
Clifford algebra. A convenient choice of
representation \cite{glw} consists of taking $\gamma^{a}$ 
to be five-dimensional 
$\gamma$-matrices for $a\ne0$ and $\gamma^0=1$. 
Note also that since they act on different
indices, $(\gamma^{a'})_i^{\ j}$ and $(\gamma^a)_{\alpha\beta}$ commute with 
each other.

It was shown in~\cite{glw} that to preserve supersymmetry 
we need only look for zero modes of
\be
\hat\delta\Theta^{\gamma'} = -{1\over 2}\epsilon^\alpha
\Gamma_\alpha^{\ \gamma'}\ ,
\ee
\ie we need to consider the off diagonal components of $\Gamma$. After some
algebra one then arrives at the following expression
\bea
\hat \delta \Theta_{\beta}^{\ j} =
&&{1\over 2}  \epsilon^{\alpha i}
\Big\{ 
{\rm det}(e^{-1})\partial_m X^{c'}{(\gamma^m)}_{\alpha\beta}
{(\gamma_{c'})}_i^{\ j}\nn
&& \hskip.75cm
- {1\over 3!} {\rm det}(e^{-1})\partial_{m_1} X^{c^\prime_1} \partial_{m_2}
X^{c^\prime_2}\partial_{m_3} X^{c^\prime_3}
(\gamma^{m_1m_2m_3})_{\alpha\beta}
{(\gamma_{c^\prime_1 c^\prime_2 c^\prime_3})}_i^{\ j}\nn
&&\hskip.75cm
+{1\over 5!}{\rm det}(e^{-1}) \partial_{m_1} X^{c^\prime_1}\dots 
\partial_{m_5} X^{c^\prime_5} (\gamma^{m_1\ldots m_5})_{\alpha\beta}
{(\gamma_{c^\prime_1 \ldots c^\prime_5})}_i^{\ j}\nn
&& \hskip.75cm
- h^{m_1m_2m_3}\partial_{m_2}X^{c_2'}\partial_{m_3}X^{c_3'}
(\gamma_{m_1})_{\alpha\beta}(\gamma_{c_2'c_3'})_i^{\ j}\nn
&& \hskip.75cm
- {1\over3}h^{m_1m_2m_3}
(\gamma_{m_1m_2m_3})_{\alpha\beta}\delta_i^{\ j}
\Big\}\ ,
\label{covsusy}
\eea
where $\gamma$-matrices always appear in tangent frame 
(\ie $\gamma_m = \delta_m^a\gamma_a$) and we use the definition
\be
\gamma^{a_1a_2a_3...a_{2n}} = 
\gamma^{[a_1}\tilde\gamma^{a_2}\gamma^{a_3}...\gamma^{a_{2n}]}\ .
\label{def}
\ee 
Clearly this agrees with the expression
given in~\cite{glw} when $h=0$. It is also easy to see, using the
self-duality of $h$, that this expression agrees with the one given 
in~\cite{hlw} when only one scalar is active.

\subsection{Travelling-waves on Calibrated Surfaces}

Before reconsidering some of the solutions discussed in section
2, we shall first discuss supersymmetric travelling wave solutions.
In the simplest setting of a flat fivebrane we expect either purely
left-moving or right-moving transverse oscillations moving
at the speed of light to preserve
supersymmetry. This is simply the fivebrane analogue of the string
travelling waves discussed
in detail in~\cite{dghw}. In fact we will be more general and
show that  such travelling waves exist on calibrated
fivebrane worldvolumes. Specifically, the waves propagate along
a flat direction 
$\s^5$ of the fivebrane and
are simply fluctuations in the 
``shape'' of the calibrated surface.

We again suppose that
we have a calibrated surface in the 
$\s^1,...,\s^4,X^7,...,X^{10}$ plane with no dependence on the
coordinate $\s^5$ and some spinor zero modes $\e$ of \p{covsusy}. 
Let us suppose that we can introduce the projector
\be
\e\gamma^0\gamma^5 = \pm\e \ ,
\ee 
and still preserve some supersymmetry. Some examples of intersecting 
fivebrane configurations for which this projector automatically follows from
the calibration are given 
in~\cite{glw}. 
To describe these waves it is helpful introduce the light-cone coordinates
\be
u = {1\over\sqrt{2}}(\s^5\mp\s^0)\ ,\quad v = {1\over\sqrt{2}}(\s^5\pm\s^0)\ ,
\ee
and from now on we assume that $a,b,c,...$ and $m,n,p,...$ take the values 
$1,2,3,4$ only. The flat metric then has the following non-zero components  
\be
\eta_{uv}=1, \quad 
\eta_{mn} = \delta_{mn}\ .
\ee
In these coordinates the projector is just $\e\gamma^{u}=0$.

Next we turn on a dependence on the coordinate 
$u$ only. To be more specific we now allow for all of the
scalars, including the scalar $X^6$, to be functions of $u$ but not
$v$, \ie $\partial_v X^{a'}=0$. The supersymmetry condition \p{covsusy}
is now, recalling that our spinors $\e$ are zero modes
when $\partial_u X^{a'}=0$,
\bea
\hat\delta\Theta =&&{1\over 2}  \epsilon
\Big\{ 
{\rm det}(e^{-1})\partial_u X^{a'}{\gamma^u}
\gamma_{a'}\nn
&& \hskip.75cm
- {1\over 3!} {\rm det}(e^{-1})\partial_{u} X^{c^\prime_1} \partial_{m_2}
X^{c^\prime_2}\partial_{m_3} X^{c^\prime_3}
\gamma^{um_2m_3}
{\gamma_{c^\prime_1 c^\prime_2 c^\prime_3}}\nn
&&\hskip.75cm
+{1\over 5!}{\rm det}(e^{-1}) \partial_{u} X^{c^\prime_1}\partial_{m_2} 
X^{c^\prime_2}\dots 
\partial_{m_5} X^{c^\prime_5} \gamma^{um_2\ldots m_5}
{\gamma_{c^\prime_1 \ldots c^\prime_5}}\Big\}\ .
\label{travelsusy}
\eea
It is not hard to see that, on account of \p{gammadecomp} and  \p{def}, 
$\gamma^{ua_1a_2} = \gamma^{u}\gamma^{a_1a_2}$ and
$\gamma^{ua_1a_2a_3a_4}=\gamma^{u}\gamma^{a_1a_2a_3a_4}$. Thus the projector
$\e\gamma^u=0$ clearly implies that
$\hat\delta\Theta =0$ and supersymmetry is again preserved for
any dependence on $u$. 

Lastly one can check that the equation of
motion, \ie the Laplacian with respect to the induced metric, 
continues to  vanish.
In addition the six-volume of the fivebrane is unaffected 
\be
{\rm det} g = -{\rm det}\bar g\ ,
\ee
where $\bar g_{mn}$ is the metric of the calibrated surface with 
$\partial_u X^{a'}=0$.
Thus these configurations are area minimising in the sense that their
six-volume is constant and is the same as the static calibrated surface,
although the spatial part of the volume form is not constant.

\subsection{Neutral Strings on Calibrated Surfaces}
 
Let us now see how one can describe neutral strings on a calibrated
surface of section 2 in the covariant formalism. 
In the $u,v$ coordinates defined above $h$ takes on the form
\be
h_{uva} = V_a\  ,\quad
h_{uab} = F_{ab}\ , \quad h_{vab} = G_{ab}\ .
\label{hform}
\ee
Self-duality then implies that $h_{abc} = \mp\epsilon_{abcd} V^d$ and also that 
$F_{ab}=\pm{1\over 2}\epsilon_{abcd}F^{cd}$  
and $G_{ab}=\mp{1\over 2}\epsilon_{abcd}G^{cd}$ 
and again $a,b,c... = 1,2,3,4$.
The 
matrix  $m$ takes the form 
\be\label{generalm}
m = \left(\matrix{1+4V^2 & -2F^2& 8V_cF^{bc}\cr
-2G^2 & 1+4V^2 & -8V_cG^{bc}\cr
-8V^cG_{ac} & 8V^cF_{ac} & (1-4V^2)\delta_a^{\ b} + 8V_aV^b\cr
}\right)\ .
\ee
To describe neutral strings we will set $V_a=0$.
If we assume that $\e$ is a preserved supersymmetry
for a calibrated surface then from \p{covsusy} we now have
\be
\hat\delta\Theta = -{1\over 2}\epsilon\Big[
F^{mn}\gamma_{vmn} +G^{mn}\gamma_{umn} 
+F^{mn}\gamma_{v}\partial_mX^{I} \partial_n X^{J}\gamma_{IJ}
+G^{mn}\gamma_{u}\partial_mX^{I} \partial_n X^{J}\gamma_{IJ}\Big]\ .
\ee
Again we suppose that we may consider the additional projector 
$\e\gamma^{u}=0$ without breaking all the supersymmetries.
Just as with \p{travelsusy} the contribution of the $F_{mn}$ terms vanishes 
automatically. However, since $\e\gamma_u \ne 0$, 
we see that
we must set $G^{mn}=0$ to preserve supersymmetry.

{}Finally we need to find the form $H$. In this case using \p{Hdef} 
one simply finds that
the only non-zero component is
\be
H_{umn} = F_{mn}\ .
\ee
Clearly the closure of $H$  asserts that  $F$ 
is an arbitrary function of $u$ and satisfies the standard
Bianchi identity. Note that  in the
Hamiltonian formalism $F$ was an arbitrary function of $\s^5$. 
{}From the 
covariant picture, we see that this
can be achieved by taking a time slice of a non-static configuration.
The most general configuration corresponds to 
an abelian instanton in the transverse space which can change its ``moduli" 
along the length of the  string as in equations \p{sol} and \p{solt}, with 
$\s^5$ replaced by $u$. Thus the left- and right-moving supersymmetric 
modes of the string live in the moduli space of abelian instantons. This 
resonates with the description of the
non-critical six-dimensional string as a sigma model on the moduli
space of non-abelian 
instantons~\cite{ABKSS}.

\subsection{Self-Dual Strings and Neutral Strings}

It is insightful to also consider the case of self-dual and neutral strings
in the covariant formalism. (We shall not consider the most general case
of adding a calibrated surface
here). In the $u,v$ coordinates defined above $h$ takes on the form
\p{hform} and we now consider $V_a\ne 0$.
As above we must set $G_{ab}=0$ in order to preserve any supersymmetry.
However the appearance of the self-dual string requires that one
of the fivebrane scalars are active, say $X = X^{6}$, and we assume that 
$\pa_{u,v}X=0$.
The condition for
supersymmetry to be preserved by this configuration is then
\be
0= \epsilon\Big[{1\over 2 }{\rm det}(e^{-1})
\gamma^m\partial_m X\gamma_6 
- \gamma^{uv}(V^m\gamma_m +{\rm det}(e^{-1})V_m\gamma^m)
-{1\over 2}\gamma^u\gamma^{mn}F_{mn} \Big]\ .
\ee
Just as was the case for a single self-dual string~\cite{hlw}, we restrict
to spinors which satisfy 
\be
\epsilon\gamma^{uv}\gamma_{6}=\epsilon\ ,
\ee
and set
\be\label{lastone}
V_a = {1\over 2} 
{1\over 1+{\rm det}(e)}\delta_a^{\ n}\partial_n X\ .
\ee
To include the neutral strings we further impose the supersymmetry projector
$\epsilon\gamma^u = 0$. 

Our next step is to calculate the three-form $H_{nmp}$ and demand that
it is closed. In the tangent frame we find
\bea
H_{uva} &=& (1+4V^2)^{-1}V_a\ ,\nn
H_{abc} &=& \mp(1-4V^2)^{-1}\epsilon_{abcd}V^d\ ,\nn
H_{uab} &=&(1-4V^2)^{-1}F_{ab}+8(1-16V^4)^{-1}(V_aV^cF_{bc}-V_bV^cF_{ac})\ ,\nn
H_{vab}&=&0\ .
\eea
However things simplify considerably when we construct $H$ in the world frame
and substitute \p{lastone}.
We find
\bea
H_{uvm} &=& {1\over4}\partial_m X \ ,\nn
H_{mnp} &=& \mp{1\over4}\epsilon_{mnpq}\delta^{qr}\partial_r X\ ,\nn
H_{umn} &=&K_{mn}\ ,\nn
H_{vmn}&=&0\ ,
\eea
where $K_{mn}=(1-4V^2)\delta^a_m\delta^b_nF_{ab}$.  
Thus the closure of $H$ leads to the simple equations
\be
\delta^{mn}\partial_m\partial_n X=0\ ,\quad
\partial_vK_{mn} =0\ ,\quad \partial_{[m}K_{np]}=0\ .
\ee
Thus $X$ is harmonic on the flat transverse space and $K_{mn}$ can
be interpreted as a (anti-) self-dual  field strength of a 
(possibly $u$-dependent) 
vector potential.

\subsection{Two Intersecting Self-Dual Strings}

As a final solution in the covariant formalism 
we now reconsider two interacting self-dual strings. 
In particular we will consider the dependence of each string on the relative 
transverse coordinates. 
The dynamics of this solution and its 
relation to the Seiberg-Witten effective action is 
discussed in~\cite{lw}.

The configuration that we are interested in may be written as 
\be
\matrix{
M5:&1&2&3&4&5& & &\cr
M2:& & & & &5&6& &\cr
M2:& & & &4& & &7&\cr
M5:&1&2&3& & &6&7&\cr}
\label{diag}
\ee
for which \p{sdsds} is a special case with only one fivebrane present.
The supersymmetry projectors for the two membranes are
\be
\epsilon\gamma^{05}\gamma_{7} = \eta\epsilon\ ,\quad
\epsilon\gamma^{04}\gamma_{6} = -\eta\epsilon\ ,
\ee
where $\eta=\pm1$. 
Note that as a result of these two projectors we may add another fivebrane
in ``for free''
\be
\epsilon\gamma^{45}\gamma_{67} = -\epsilon\ ,
\ee 
which we have already included in \p{diag}. Thus we expect to 
obtain 
two self-dual strings (or a single non-trivial string) on
a Riemann surface corresponding to \p{riemann}.

Rather than use the light
cone coordinates of the previous sections it
is helpful instead to introduce complex notation 
\be
z = \s^4+i \s^5\ ,\quad
s = X^6 +iX^7\ ,
\ee
with the derivatives 
$\partial = \partial_z$, $\bar \partial=\partial_{\bar z}$.
The projectors can then simply be written as 
\be
\epsilon\gamma_{0z}= \eta \epsilon \gamma_{\bar s}\ ,\quad
\epsilon\gamma^z\gamma_s =0\ .
\label{q}
\ee
In total this configuration preserves one quarter of the  fivebrane's
worldvolume supersymmetry. In this section the indices $a,b,c,...=0,1,2,3$ 
are in tangent
frame, the indices $\mu,...=0,1,2,3$ are in world frame and 
we also take $i,j,k,...=1,2,3$ in the world frame. 
For simplicity 
we will only consider the solution to 
second order in the spatial derivatives $\partial_i$. 
This will simplify our calculations,
however it is reasonable to hope that the end result is 
valid to all orders.

Our next step is to  decompose $h$ into a four-dimensional
vector $V_a$ and anti-symmetric tensor ${\cal F}_{ab}$ 
as follows (all indices are
in the tangent frame)
\be
h_{abz} = {\cal F}_{ab}\ , \quad 
h_{ab\bz}={\bar {\cal F}}_{ab}\ ,\quad h_{az\bz} = iV_a\ .
\label{hdecomp}
\ee
Self-duality implies that $h_{abc} = 2\epsilon_{abcd}V^d$ and
${\cal F}_{ab} = {i\over2}\epsilon_{abcd}{\cal F}^{cd}$. 
{}For the convenience of the reader we list the components of the 
vielbein $e_{m}^{\ a}$ for the geometry resulting from $s$
\bea
e_{\mu}^{\ a} & =& \delta_{\mu}^{\ a} -{1\over2} 
\left(1\over \det e\right)^2  \left(
\bar\partial s\partial s\partial_{\mu}\bar s\partial^a\bar s 
+\partial \bar s\bar\partial\bar s\partial_{\mu}s\partial^a s\right)\ ,\nn
&&\hskip1.1cm+{1\over4}\left({1+\ds+\dsb\over (\det e)^2}\right)
\left(\partial_{\mu}s\partial^{a}\bar s+\partial_{\mu}\bar s\partial^a s\right)
\ ,\nn
e_{\mu}^{\ z} &=& {(X^2-\ds)\bar\partial s\partial_{\mu}\bar s
+ (X^2-\dsb)\bar\partial\bar s\partial_{\mu}s\over X \det e}\ ,\nn
e_{\mu}^{\ \bz} &=& {(X^2-\ds)\partial \bar s\partial_{\mu}s
+ (X^2-\dsb)\partial s\partial_{\mu}\bar s\over X \det e}\ ,\nn
e_z^{\ \bz} &=& {\partial s\partial \bar s\over X}\ ,\quad\quad
e_{\bz}^{\ z} =  {\bar\partial \bar s\bar\partial s\over X}\ ,\nn
e_z^{\ z}&=&
e_{\bz}^{\ \bz} = X\ ,\nn
\eea
where
\bea
X^2 &=& {1\over2}\left[(1+\ds+\dsb) + \det e\right]\ ,\nn
\det e  &=& \sqrt{(1+\ds+\dsb)^2 - 4\ds\dsb}\ .\nn
\eea

Next we note that the 
projectors \p{q} imply that there are four 
independent terms appearing in supersymmetry condition
$\hat\delta\Theta=0$ proportional to
\be
\epsilon\gamma_{0iz}\ ,\quad
\epsilon\gamma_{0z\bar z}\ , \quad 
\epsilon\gamma_{iz\bar z}\ , \quad
\epsilon\gamma_0\ , 
\ee
and their complex conjugates.
Thus we may obtain the  Bogomol'nyi equations by setting the corresponding
coefficients to zero. Using the decomposition \p{hdecomp} this yields
\bea
{\cal F}_{0i} 
&=& {1\over8}\eta\left({1+\ds-\dsb\over X^2 - \dsb}\right)
\left(
{X^2\partial_is + \partial\bar s\partial s\partial_i\bar s\over X \det e}
\right)\ ,\nn
V_0 &=& +{i\over16}\eta\left({1+\ds-\dsb\over (X^2-\dsb)^2}\right)\left[
(1+\ds+\dsb){\bar\partial s\partial_is\partial^i\bar s\over(\det e)^2}\right. 
\nn
&&\left.\quad\quad\quad +\dsb
{(\partial s\partial_i\bar s\partial^i\bar s-\bar\partial\bar 
s\partial_i\partial^is)
\over (\det e)^2}\right]+ {i\over4}\eta{\bar\partial s\over X^2 - \dsb},\nn
V_i &=& {1\over16}\eta\bar\partial s\left({1+\ds-\dsb\over (X^2-\dsb)^2}\right)
{\epsilon_{ijk}\partial^js\partial^k\bar s\over \det e}\ ,\nn
{\bar \partial}s &=& -\partial\bar s\ ,
\eea
respectively. 
Here all indices are raised and lowered with the flat metric.

The next step is to calculate the physical three-form $H$ defined in 
\p{Hdef} which most
naturally appears in the equations of motion. To help the reader we  
give here the matrix $m^{-1}=Q^{-1}(1+2k)$ 
\bea
m^{-1} = Q^{-1}\left(\matrix{
\delta_{\mu}^{\ \nu}+ 2k_{\mu}^{\ \nu}
&32i\bar\kappa v_0{\bar {\cal F}}_{\mu}^{\ 0}
&-32i\kappa v_0{{\cal F}}_{\mu}^{\ 0}
\cr
-16i\kappa v_0{{\cal F}}^{\nu 0}
&1-16v_0^2
&4\kappa^2{\cal F}^2
\cr
16i\bar\kappa v_0{\bar{\cal F}}^{\nu 0}
&4\bar\kappa^2{\bar{\cal F}}^2
&1-16v_0^2\cr
}\right)\ ,
\eea
where
\bea
k_{\mu}^{\ \nu}&=&8v_0^2\delta_\mu^{\ \nu} + 16v_\mu v^\nu 
+4|\kappa|^2{\cal F}_{\mu\lambda}{\bar {\cal F}}^{\nu\lambda} 
+ 4|\kappa|^2{\bar {\cal F}}_{\mu\lambda}{\cal F}^{\nu\lambda}\ ,\nn
Q&=& 1 - 256v_0^2(v_0^2 - 2|\kappa|^2{\cal F}_{0i}{\bar {\cal F}}^{0i})\ .\nn
\eea
Despite the complicated form of these Bogomol'nyi equations one finds 
after a lengthy calculation that
the three-form $H$ takes on a relatively simple form. In particular, in 
the world frame 
\bea
H_{iz\bz} &=&0 \ ,\nn 
H_{0ij} &=& 0\ ,\nn
H_{0iz} &=&  {1\over 8}\eta\partial_i s\ ,\quad \quad
H_{0i\bz} = {1\over 8}\eta\partial_i\bar s\ ,\nn
H_{ijz} &=&  {i\over 8}\eta\epsilon_{ijk}\partial^ks\ ,\quad\quad
H_{ij\bz} =  -{i\over 8}\eta\epsilon_{ijk}\partial^k\bar s \ ,\nn
H_{0z\bz} &=& -{1\over 4}\eta {\bar \partial}s\ , \nn
H_{ijk} &=& -{i\over 8}\eta{\epsilon_{ijk}}\left({
4{\bar \partial}s+2{\bar \partial}s\partial_i s\partial^i\bar s
+\partial s\partial_i\bar s\partial^i\bar s
-{\bar \partial}\bar s\partial_is\partial^is \over 1+\ds-\dsb}
\right)\ .\nn
\eea
Note that $s$ is not holomorphic but instead satisfies 
${\bar \partial}s = -\partial\bar s$.
Indeed one sees that the complete 
dependence of fields on the relative transverse coordinates
of the two self-dual strings is given by the non-holomorphicity of $s$.
The  equation of motion for $s$ can then be found 
by demanding that $dH=0$. If we set $\partial s=\bar \partial s=0$ we then
arrive at the solution in section five, with the equation of motion
$\partial^i\partial_i s=0$ resulting from $\partial_{[i}H_{jkz]}=0$.

\section{Conclusion}

In this paper we have examined the conditions for the preservation of
the non-linear supersymmetry of the fivebrane for any bosonic configuration
in both the Hamiltonian and covariant formalisms. Furthermore we formulated 
the conditions and field equations for
several supersymmetric solitons with a non-zero three-form field. 
In particular we found that self-dual and neutral strings correspond to  
harmonic functions and
instantons on calibrated surfaces, respectively. 
To produce a specific solution one typically needs to solve the field
equations, \ie construct harmonic functions or  instantons
on calibrated surfaces, and it would be interesting to consider 
some specific cases in more detail.
All of the cases we have considered (perhaps setting $\partial_5 =0$) may be 
dimensionally reduced along $\s^5$ to obtain solutions of the D4-brane 
worldvolume theory. We expect that they have a natural generalisation to
the non-Abelian DBI theory and this would be worth checking. 

We have
by no means produced an exhaustive list of solutions however it is hoped
that the reader will have gained some insight into the general form of the
solutions  by means of these examples. In addition we hope that the
reader will have gained some further understanding of the complicated
non-linear theory on the fivebrane worldvolume in both the formalisms
discussed.

\vskip 0.5cm
\noindent
{\bf Acknowledgements}:
JPG is supported in part by the EPSRC and thanks Jeff Harvey and 
the Enrico Fermi Institute, 
University of Chicago for hospitality during the completion
of this work.
 

\end{document}